\documentclass[pre,aps,twocolumn,amsmath,amssymb,floatfix]{revtex4}
\usepackage{graphicx}
\usepackage{amsfonts}
\usepackage{epsfig}
\usepackage{dcolumn}
\usepackage{amsthm}
\usepackage{color}

\usepackage{latexsym}
\usepackage{amscd}
\usepackage{hyperref}
\usepackage[mathscr]{eucal}
 
\usepackage{bm}      
\usepackage{ifpdf}
\usepackage{bbm}
\usepackage{float}
\usepackage{xcolor}
\usepackage[fleqn]{mathtools}
\begin{document}

\newcommand{\newc}{\newcommand}

\newc{\beq}{\begin{equation}}
\newc{\eeq}{\begin{equation}}
\newc{\ovl}{\overline}
\newc{\bc}{\begin{center}}
\newc{\ec}{\end{center}}
\newc{\tr}{\mbox{tr}}
\newc{\pd}{\partial}
\newc{\dqv}{\delta\vec{q}}
\newc{\dpv}{\delta\vec{p}}
 \newc{\f}{\frac} 
 \title{Universal scaling of higher-order spacing ratios in Gaussian random matrices}
\author{Udaysinh T. Bhosale}
\email{udaysinhbhosale@phy.vnit.ac.in}
\affiliation{Department of Physics, Visvesvaraya National Institute of Technology, Nagpur 440010, India}
\date{\today}
\begin{abstract}

%

Higher-order spacing ratios are investigated analytically using a Wigner-like surmise for Gaussian ensembles of random matrices.
For $k$-th order spacing ratio $(r^{(k)}$, $k>1)$ the matrix of dimension $2k+1$ is considered.
A universal scaling relation for this ratio, known from earlier numerical studies,
is proved in the asymptotic limits of $r^{(k)}\rightarrow0$ and $r^{(k)}\rightarrow \infty$.


\end{abstract}
 
\maketitle
 
\section{Introduction}
  
Random matrix theory (RMT), introduced for more than fifty years, has been applied successfully in various fields 
\cite{Mehtabook,akemann2011oxford,Forresterbook}. Originally it was introduced to explain intricate spectra of heavy nucleus 
\cite{PorterBook}. Later, it has found applications in complex networks \cite{jalan2007random,rai2015application}, many-body physics 
\cite{hutchinson2015random,wells2014quantum,rao2020higher, rao2021critical,rao2022random}, wireless communications
\cite{tulino2004random}, etc. One of the main objectives of RMT is to study the spectral fluctuations in these systems. These 
fluctuations can be used to characterize the different types of phases of these complex systems. For example, 
integrable to chaotic limits of the underlying classical systems \cite{Bohigas84,Hakke87,Reichlbook},
thermal or localized phases of condensed matter systems \cite{rao2020higher,rao2021critical,rao2022random,Oganesyan2007}, etc.
Bohigas, Giannoni, and Schmit conjectured that the eigenvalue fluctuations in a quantum chaotic system can be modelled by one of 
the three classical ensembles of RMT depending on the underlying symmetry. These ensembles having Dyson indices as $\beta=1$, $2$ 
and $4$ respectively corresponds to Hermitian random matrices whose entries are chosen/distributed independently, respectively, 
as real (GOE), complex (GUE), or quaternionic (GSE) random variables  \cite{Mehtabook}.

The most popular measure to model the spectral fluctuations is the nearest neighbour (NN) level spacings, $s_i=E_{i+1}-E_i$, 
where $E_i$, $i=1,2,\ldots$ are the eigenvalues of the given Hamiltonian $H$. 
A surmise by Wigner states that in a time-reversal invariant 
system ($\beta=1$) which do not have a spin degree of freedom, these spacings are distributed as $P(s)=(\pi/2)s\exp(-\pi s^2/4)$, 
which indicates the level repulsion. This result is very close to the exact one which has been obtained later on 
\cite{dietz1990taylor,Mehtabook,Forresterbook}.
For such systems, Gaussian Orthogonal Ensemble (GOE) is well suited to study the statistical properties of their spectra. There are 
other ensembles also commonly used in RMT, namely, Gaussian unitary ensemble (GUE) and Gaussian symplectic ensemble (GSE) having 
Dyson index $\beta=2$ and $4$ respectively. The GUE is applicable systems systems without time reversal whereas GSE to spin $1/2$ 
systems having time reversal respectively but no rotational symmetry \cite{Mehtabook,Forresterbook}. The member matrices of these 
families are real symmetric, complex Hermitian, and quaternion self-dual, respectively \cite{dumitriu2002matrix}. These ensembles 
have been implemented successfully in various fields \cite{akemann2011oxford,cotler2017black}. In this paper, the Gaussian 
ensembles are studied in detail and various analytical results are obtained.

When the fluctuations are studied using the spacing distribution, one needs to carry out the procedure called  unfolding the spectra 
which takes off the system-dependent spectral properties, i.e., the average part of the density of states (DOS) 
\cite{Mehtabook,PorterBook,Haakebook,BruusHenrik1997,Berry77a,prosen1993energy,Guhr98}. Thus, the comprehension of the system's 
DOS is required. This procedure is nontrivial and cumbersome especially in many-body physics where not enough 
eigenvalues are available to get a good fit of the DOS 
\cite{BruusHenrik1997,plerou2002random,Haakebook,VadimOganesyan2009,GomezMisleading2002}. It can reduce the accuracy of statistical 
tests in such 
systems. It is also shown that short-range correlations are not sensitive to the unfolding method whereas the long-range level 
correlations are strongly dependent on the unfolding procedure employed (see Ref.\cite{GomezMisleading2002} for more details).


This challenging problem can be resolved by using the NN spacing ratios \cite{VadimOganesyan2007},
because it is independent of the local DOS which implies that unfolding is not required.
It is defined as $r_i=s_{i+1}/s_{i}$, $i=1$, $2$, $\ldots$.
For the case of Gaussian ensembles, a Wigner-like surmise for the distribution of $r_i$, $P(r)$ has been obtained as follows 
\cite{BogomolnyDistribution2013,BogomolnyJoint2013}:
\begin{equation}
\label{Eq:PRBeta}
 P(r,\beta)=\frac{1}{Z_{\beta}} \frac{(r+r^2)^\beta}{(1+r+r^2)^{(1+3\beta/2)}},\,\,\,\,\beta=1,2,4
\end{equation}
where $Z_{\beta}$ is the normalization constant. 
It must be noted here that this distribution has been derived using {\it only} three eigenvalues with the Gaussian weight.
The expression will change with the matrix dimensions $N$ (as observed in Ref.\cite{BogomolnyJoint2013} for $N=4$) 
as well as the weight. Although small deviations for smaller $N$ are observed and pointed out in the Ref.\cite{BogomolnyDistribution2013}, 
this works as a very good approximation for large $N$ and in the bulk of the spectrum. The exact analytical expression
for any $N$ still remains an open question.

This distribution has found many applications, to study eigenvalue statistics in spin systems 
\cite{buijsman2018random,VadimOganesyan2007,VadimOganesyan2009,ArijeetPal2010,ShankarIyer2013,cuevas2012level,biroli2012difference,rao2022random,kundu2022spectral},
in triangular billiards \cite{lozej2022quantum}, in the Hessians of artificial neural networks \cite{baskerville2022appearance},
in Sachdev-Ye-Kitaev model \cite{FadiSunPeriodic2020,FadiSunClassification2020,nosaka2020quantum,fremling2022bipartite,sa2022q}, 
in quantum field theory \cite{srdinvsek2021signatures},
to quantify symmetries in various complex systems \cite{tekur2018symmetry,bhosale2021superposition}.

As pointed in Refs.\cite{BogomolnyDistribution2013,BogomolnyJoint2013}, the distributions in the Eq.~(\ref{Eq:PRBeta}) are universal, 
i.e. they can be applied without any unfolding or renormalization to the eigenvalues of complex physical systems. 
%
It also shows an interesting behavior (thus, universal) as follows:
\begin{eqnarray}
\begin{split}
 &P(r,\beta) \rightarrow r^{\beta} \;\;\; \;\;\; \;\;\;  \mbox{for}\;\;\; r \rightarrow 0\\
 &P(r,\beta) \rightarrow  r^{-2-\beta} \;\;\; \mbox{for} \;\;\; r \rightarrow \infty. \\
 \end{split}
\label{Eq:Universal1}
\end{eqnarray}
A correction ansatz, $\delta P_{\mbox{fit}}(r)$, was given to the Eq.~(\ref{Eq:PRBeta}) such that 
$P(r)+\delta P_{\mbox{fit}}(r)$ fits very well for all values of $N$ \cite{BogomolnyDistribution2013}, where
\begin{eqnarray}\label{Eq:CorrectionTerm}
\delta P_{\mbox{fit}}(r)=\dfrac{C}{(1+r)^2}\left[\left(r+\dfrac{1}{r}\right)^{-\beta}-c_{\beta}\left(r+\dfrac{1}{r}\right)^{-1-\beta}\right].
\end{eqnarray}
Here, $C$ and $c_{\beta}$ are some constants. And it can be seen that despite this correction term the universal behaviour remain 
unchanged in the Eq.~(\ref{Eq:Universal1}). 

Variants of these spacings are proposed and applied to various systems 
\cite{BogomolnyJoint2013,chavda2014poisson,kota2018embedded,HarshiniExact2018} which includes the 
generalization to the complex eigenvalues 
\cite{ComplexProsen2020,dusa2022approximation,luo2021universality,li2021spectral,sa2020spectral,prasad2022dissipative}.

In this paper, we study the non-overlapping $k$-th order spacing ratios, which are defined such that no eigenvalue is common between the 
spacings in the numerator and denominator. 
It is defined as follows:
\begin{equation}
 r_{i}^{(k)}=\frac{s_{i+k}^{(k)}}{s_{i}^{(k)}}=\frac{E_{i+2k}-E_{i+k}}{E_{i+k}-E_{i}},\;\;\; i,k=1,2,3,\ldots.
\end{equation}
The case $k=1$ corresponds to the earlier solved case from the Ref.\cite{BogomolnyDistribution2013}.
Its distribution has found applications to study higher-order fluctuation statistics in the Gaussian \cite{Harshini2018a}, 
Wishart \cite{UdaysinhBhosaleScaling2018} and circular ensembles \cite{Harshini2018a}. 
 An important scaling relation in these 
cases, in the asymptotic limit of $N\rightarrow \infty$ and in bulk of the spectra, by extensive numerical 
computations is given as follows \cite{Harshini2018a,UdaysinhBhosaleScaling2018}:
\begin{eqnarray}
\begin{split}
P^{k}(r,\beta)&=P(r,\beta'),\,\,\,\,\beta\geq 1\\
\beta'&=\frac{k(k+1)}{2}\;\beta + (k-1),\,\,\,\,k \geq 1.
\end{split}
\label{Eq:HigherOrder}
\end{eqnarray}
It means, the distribution of $k$-th order spacing ratio for a given $\beta$ ensemble is same as that of NN spacing ratios of some 
other ensemble with Dyson index $\beta'(>\beta)$. It should be noted that the exact analytical expression for any $k\geq2$ and any 
$N$ is not known yet but the numerics suggest that the Eq.~(\ref{Eq:HigherOrder}) works very well for large enough $N$ 
and in the bulk of the spectra. For given $k$, the effect of increasing $N$ is studied numerically
in Ref.\cite{Harshini2018a}. There it is shown that for given $k$, however large, the fitted $\beta'$ converges to the value given 
in the Eq.~(\ref{Eq:HigherOrder}) as $N$ is increased. For smaller $N$, we expect the same expression 
in Eq.~(\ref{Eq:CorrectionTerm}) can be used as the correction term but with modified index $\beta'$.
For this, we have assumed that for large $N$ the asymptotic behavior for small and large $r$ 
is same for both $P(r,\beta')$ and $\delta P_{\mbox{fit}}(r,\beta')$ for $k \geq 2$ \cite{BogomolnyDistribution2013}.
It should be noted that Eqs. \ref{Eq:PRBeta} and \ref{Eq:CorrectionTerm} taken together still represent an approximation.
Thus, it is more likely that the exact (currently unknown) expression for the $k$-th spacing distribution also shares the 
same asymptotics of the Eq.(\ref{Eq:HigherOrder}).

This relation has been employed successfully to various
physical systems like chaotic billiards, Floquet systems, circular ensembles, spin chains, observed stock market, etc.
\cite{Harshini2018a,UdaysinhBhosaleScaling2018,rao2020distribution,buijsman2018random,rao2020higher,de2021intermediate,sierant2020model}
, to estimate the number of symmetries in complex physical systems \cite{tekur2018symmetry,bhosale2021superposition}.
It should be noted that, a similar scaling relation between the higher-order and NN {\it spacing distributions} has been proposed 
earlier in Refs.\cite{kahn1963statistical,AbulMagd1999}, later proved partly in Ref.\cite{UdaysinhBhosaleScaling2018} and 
completely in Ref.\cite{rao2020higher} using a Wigner-like surmise for the Gaussian ensembles. 
It is shown numerically in the Ref.\cite{rao2020higher}; using  random spin systems, nontrivial zeros of the Riemann $\zeta$ 
function and Gaussian ensemble; that as $N$ is increased the deviations from the surmise become smaller and smaller. Although 
the bulk statistics, for given $\beta$, is same in these three ensembles (Gaussian+Wishart+circular) in the large-$N$ limit, 
the physical systems described by them are very different from each other \cite{akemann2011oxford}.


It should be noted that the result in Eq.~(\ref{Eq:HigherOrder}) for the spacing ratios is a purely numerical one except for few special cases 
\cite{MehtaDyson63,forrester2004correlations,forrester2009random}. Thus, a complete analytical understanding of this result is 
lacking. In this paper, we give partial analytical support to it since proving the entire result is mathematically challenging. 
If this result is correct then using the universality aspect as per Eq.~(\ref{Eq:Universal1})
one can conclude that, for the higher-order spacing ratios, following must be true:
\begin{eqnarray}
\begin{split}
 &P^{k}(r,\beta)=P(r,\beta')   \rightarrow r^{\beta'} \;\;\; \;\;\; \;\;\;  \mbox{for}\;\;\; r \rightarrow 0\\
 &P^{k}(r,\beta)=P(r,\beta')  \rightarrow  r^{-2-\beta'} \;\;\; \mbox{for} \;\;\; r \rightarrow \infty, \\
 \end{split}
\label{Eq:Universal2}
\end{eqnarray}
with $\beta'$ as per the Eq.~(\ref{Eq:HigherOrder}).
In this paper, we derive analytically Eq.~(\ref{Eq:Universal2}) using Wigner-like surmise for the Gaussian ensembles.

The structure of the paper is as follows: In Sec.~\ref{sec:ResultsKis2} we present the results for the case $k=2$. 
In Sec.~\ref{sec:ResultsKisGeneral} (Sec.~\ref{Sec:ResultsRinInfinity}) the general result for any $k$ is 
provided in the limit $r \rightarrow 0$ ($r \rightarrow \infty$).
In Sec.~\ref{Sec:UncorrelatedSpectra} the case of uncorrelated spectra is studied in the asymptotic limits and related to the 
results from the Gaussian ensembles. Finally, in Sec.~\ref{Sec:Summary} a summary of the results and conclusions are presented.





\begin{widetext}
\section{Results: $k=2$ Case}
\label{sec:ResultsKis2}

Before entering into the main results we would like to mention that throughout the paper our calculations are restricted to the 
simplest and lowest matrix dimensions $N$ such that for given order $k$, $N=2k+1$ (which is our Wigner-like surmise). 
One should note that, in order to study $k$-order spacing ratios one should have at least $2k+1$ levels to be in the RMT regime. 
In a Hamiltonian system these levels $E_i$ become eigenenergies. Earlier studies from the 
Refs.\cite{osborn1998thouless} indicates that the difference $E_{2k+1}-E_1$ should be less than
the systems Thouless energy ($E_c$) for RMT to hold true. This is an important point to be noted when applying our results 
to various physical systems \cite{akemann2011oxford}.
Let us first start with the joint probability distribution function (jpdf) of the Gaussian ensemble which is given as follows:
\begin{equation}\label{Eq:jpdfGaussian}
 f(\{E_l\}) \propto \prod_{1\leq i<j\leq N}^{} |E_i-E_j|^{\beta}  \exp\left(-A \sum_{i=1}^{N} E_i^2 \right),
\end{equation}
where $\beta=1$, $2$ and $4$ for GOE, GUE, and GSE, respectively \cite{Mehtabook,Forresterbook}. Without loss of generalities, 
we will be assuming $E_1\leq E_2 \leq \ldots \leq E_N$ throughout this paper. Firstly, consider the case of $k=2$ and general $\beta$. Here, 
for Wigner-like surmise, we need to have five eigenvalues \cite{rao2020higher}. Then we get:
\begin{eqnarray}
 r^{(2)}=\dfrac{E_5-E_3}{E_3-E_1}.
\end{eqnarray}
Then the distribution $P\left(r^{(2)}\right)$ becomes \cite{rao2020higher}:
\begin{eqnarray}
P\left(r^{(2)}\right) \propto   \int_{-\infty}^\infty \cdots \int_{-\infty}^\infty \prod_{1\leq i<j \leq 5}^{} |E_i-E_j|^{\beta} 
\exp\left(-A \sum_{i=1}^{5} E_i^2 \right) \delta\left(r^{(2)}-\dfrac{E_5-E_3}{E_3-E_1}\right)  \prod_{i=1}^{5} dE_i.
\end{eqnarray}
We first change the variables to $x_i=E_{i+1}-E_{i}$ for $i=1$ to $4$
and $x_5=\sum_{i=1}^5 E_i$
\cite{rao2020higher}.
Then $P\left(r^{(2)}\right)$ simplifies to \cite{rao2020higher}:
\begin{eqnarray}
\begin{split}
 P\left(r^{(2)}\right) \propto &  \int_{0}^\infty \cdots \int_{0}^\infty
  \dfrac{\partial(E_1,\ldots,E_{5})}{\partial(x_1,\ldots,x_{5})}\nonumber
\left( \prod_{i=1}^4 \prod_{j=i}^4 \left|\sum_{l=i}^j x_l\right|^{\beta} \right)
\exp\left\{-\dfrac{A}{5} \left[\sum_{i=1}^{4}  \sum_{j=i}^{4}  \left(\sum_{l=i}^{j} x_l \right)^2 + x_5^2\right] \right\}\\
&\delta\left(r^{(2)}-\dfrac{x_3+x_4}{x_1+x_2}\right) \prod_{i=1}^{5} dx_i.
\end{split}
\end{eqnarray}
Here, the Jacobian $\dfrac{\partial(E_1,\ldots,E_{5})}{\partial(x_1,\ldots,x_{5})}$ and integral for $x_5$ are constants 
that can be absorbed into the normalization factor, and using the property of the delta function we obtain:
\begin{eqnarray}\label{Eq:kis2Integral}
\begin{split}
P\left(r^{(2)}\right) \propto  \int_{0}^\infty \cdots \int_{0}^\infty&(x_1+x_2) 
\left( \prod_{i=1}^4 \prod_{j=i}^4 \left|\sum_{l=i}^j x_l\right|^{\beta} \right)
\exp\left\{-\dfrac{A}{5} \left[    \sum_{i=1}^{4}  \sum_{j=i}^{4}  \left(\sum_{l=i}^{j} x_l \right)^2 \right] \right\}\\
&\delta\left(r^{(2)}(x_1+x_2)-(x_3+x_4)\right) \prod_{i=1}^{4} dx_i.\label{Eq:Integral2}
\end{split}
\end{eqnarray}

First the integral over $x_4$ in the Eq.~(\ref{Eq:kis2Integral}) is carried out. 
Then, $x_4$ will be replaced by $r(x_1+x_2)-x_3$ due to the delta function in Eq.~(\ref{Eq:Integral2}) 
(here, we define $r^{(2)}=r$ for simplicity of the notation). And then the limits of 
integration of $x_3$ will be from $0$ to $r(x_1+x_2)$. 
Thus, our strategy is, first to find the lowest degree polynomial in $x_3$ and $x_4$, since in the limit $r\rightarrow 0$
the integration over both $x_3$ and $x_4$ will give us the leading order term in $r$.

%
Thus, first consider the following term from the integrand of Eq.~(\ref{Eq:Integral2}):
\begin{eqnarray}
 \prod_{i=1}^4 \prod_{j=i}^4 \left|\sum_{l=i}^j x_l\right|^{\beta}, 
\end{eqnarray}
which can be expanded to: 
\begin{eqnarray}
\label{Eq:Term1}
&&\left\{x_1 x_2 x_3 x_4 (x_1+x_2) (x_1+x_2+x_3) (x_1+x_2+x_3+x_4) (x_2+x_3)  (x_2+x_3+x_4) (x_3+x_4)\right\}^{\beta}\\
&=&\left\{ x_3 x_4 (x_3+x_4)\times x_1 x_2 (x_1+x_2) (x_1+x_2+x_3) (x_1+x_2+x_3+x_4) (x_2+x_3)  (x_2+x_3+x_4) \right\}^{\beta}
\end{eqnarray}
This can be written as 
\begin{eqnarray}\label{Eq:x3x4}
\left\{x_3 x_4 (x_3+x_4) \left[f_1(x_1,x_2)+f_2(x_1,x_2,x_3,x_4)\right]  \right\}^{\beta},
\end{eqnarray}
where $f_1$ and $f_2$ are polynomial functions of the respective arguments.
This kind of split is possible because apart from the term $x_3 x_4 (x_3+x_4)$
all the terms contain at least one variable from $x_1$ and $x_2$.
%
%
%
%
The exact form of $f_i$'s can be found easily but are not required for our purpose here. That's because if we see the 
Eq.(\ref{Eq:x3x4}) carefully, after expanding it, the lowest order polynomial in $x_3$ and $x_4$ turns out to be 
$\left\{x_3 x_4 (x_3+x_4) \right\}^{\beta}   f_1^{\beta}(x_1,x_2) $ with order $3\beta$ in $r$. And it is this term 
which will give us the required lowest  power of $r$ in the limit $r \rightarrow 0$. This will be clear in the subsequent 
calculations. Thus, the Eq.~(\ref{Eq:Integral2}) becomes:
\begin{eqnarray}\label{}
\begin{split}
P\left(r^{(2)}\right) \propto &  \int_{0}^\infty \cdots \int_{0}^\infty (x_1+x_2) 
\left\{x_3 x_4 (x_3+x_4) \left[f_1(x_1,x_2)+f_2(x_1,x_2,x_3,x_4)\right]  \right\}^{\beta}\\
&\exp\left\{-\dfrac{A}{5} \left[    \sum_{i=1}^{4}  \sum_{j=i}^{4}  \left(\sum_{l=i}^{j} x_l \right)^2 \right] \right\}
\delta\left(r^{(2)}(x_1+x_2)-(x_3+x_4)\right) \prod_{i=1}^{4} dx_i. 
\end{split}
\end{eqnarray}

 
 Integrating over $x_4$ and simplifying further we get:
 \begin{eqnarray}\label{}
\begin{split}
 P\left(r^{(2)}\right) \propto  \iint_{x_1,x_2=0}^{\infty}  \int_{x_3=0}^{r (x_1+x_2)} & (x_1+x_2) 
\left\{x_3  (rx_1+rx_2-x_3) r(x_1+x_2) \left[f_1(x_1,x_2)+f_2(x_1,x_2,x_3,rx_1+rx_2-x_3)\right]  \right\}^{\beta}\\
&\exp\left\{-\dfrac{A}{5} \left[  (2 \left(2+r+2 r^2\right) {x_1}^2+\left(6+4 r+4 r^2\right)  {x_2}^2
+ 2{x_1}{x_2}\left(3+3r+4r^2\right)  \right] \right\}\\
&\exp\left\{-\dfrac{A}{5}\left[ 4 {x_3}^2 + {x_3}\left((4-2r){x_2}+(2-2r){x_1}\right)\right]\right\}\prod_{i=1}^{3} dx_i. 
\end{split}
\end{eqnarray}
This we write as follows:
\begin{eqnarray}\label{Eq:Pr2One}
\begin{split}
 P\left(r^{(2)}\right) \propto \iint_{x_1,x_2=0}^{\infty}  &  I_{x_3}(x_1,x_2,r)
  \exp\left\{-\dfrac{A}{5} \left[  (2 \left(2+r+2 r^2\right) {x_1}^2+\left(6+4 r+4 r^2\right)  {x_2}^2
+ 2{x_1}{x_2}\left(3+3r+4r^2\right)  \right] \right\}\\
 &(x_1+x_2)^{1+\beta} \prod_{i=1}^{2} dx_i,
 \end{split}
\end{eqnarray}
where the $x_3$-integral is given as follows:
\begin{eqnarray}
\begin{split}
 I_{x_3}(x_1,x_2,r)=\int_{x_3=0}^{r (x_1+x_2)} &
 \left\{x_3  (rx_1+rx_2-x_3) r  \left[f_1(x_1,x_2)+f_2(x_1,x_2,x_3,rx_1+rx_2-x_3)\right]  \right\}^{\beta}\\
& \exp\left\{-\dfrac{A}{5}\left[ 4 {x_3}^2 + {x_3}\left((4-2r){x_2}+(2-2r){x_1}\right)\right]\right\} dx_3.
\end{split}
\end{eqnarray}

Here, we are interested only to find the leading order term in $r$ in the limit $r \rightarrow 0$ of $P\left(r^{(2)}\right)$ 
and thus to find the dominant term in $r$. It can be seen from the Eq.(\ref{Eq:Pr2One}) that the leading order will 
come only from that of $I_{x_3}(x_1,x_2,r)$. Whereas integration over $x_1$ and $x_2$ are converging and will give another constant,
keeping the exponent of $r$ unchanged. Thus, we need to find only the lowest power 
of $r$. Using the fact that the limit and the integral can be interchanged \cite{kamihigashi2020interchanging}, and the limit of the 
product is the product of the limits, let us first consider the term $I_{x_3}(x_1,x_2,r)$. 
Let us first consider the term $(rx_1+rx_2-x_3)^{\beta}$. It can be simplified as follows:
\begin{eqnarray}
 (rx_1+rx_2-x_3)^{\beta}=\sum_{q=0}^{\beta} {\beta\choose q} r^{q} (x_1+x_2)^{q} (-x_3)^{\beta-q}
\end{eqnarray}
Thus, 
\begin{eqnarray}\label{Eq:Ix3One}
\begin{split}
 I_{x_3}(x_1,x_2,r)=\int_{x_3=0}^{r (x_1+x_2)} & \left[x_3\right]^{\beta}   
 \left[ \sum_{q=0}^{\beta} {\beta\choose q} r^{q} (x_1+x_2)^{q} (-x_3)^{\beta-q} \right]^{\beta}  \left[r \right]
 \left[f_1(x_1,x_2)+f_2(x_1,x_2,x_3,rx_1+rx_2-x_3)\right]^{\beta} \\
& \exp\left\{-\dfrac{A}{5}\left[ 4 {x_3}^2 + {x_3}\left((4-2r){x_2}+(2-2r){x_1}\right)\right]\right\} dx_3.
\end{split}
\end{eqnarray}
The square brackets around various terms are put in order to address them individually. Now, our strategy is to 
find lowest order of the polynomial in $x_3$ and $r$. Then we will use the Eq.(\ref{Eq:IntegralIdentity1}) given as 
follows: 
\begin{eqnarray}\label{Eq:IntegralIdentity1}
  \int\limits_{y=0}^{a}   y^p dy \;\propto \;a^{p+1},
\end{eqnarray}
and evaluate the integral.
The first square bracket will give exponent of $\beta$ in $r$ for $x_3$,  second square bracket will give $\beta-q$,
fourth square bracket and the exponential term will give $0$ as the lowest exponent of $x_3$.
The second and third square bracket together will give $q+\beta$ as exponent of $r$.
Thus, using Eq.(\ref{Eq:IntegralIdentity1}) in the Eq.(\ref{Eq:Ix3One}), the leading term in $r$ in the 
$I_{x_3}(x_1,x_2,r)$ and eventually in $P\left(r^{(2)}\right)$ is $ r^{3\beta+1}$. The extra `+1' factor in the 
exponent comes from the integration measure $dx_3$.
Thus, we obtain that $P\left(r^{(2)}\right) \rightarrow r^{3\beta+1}$ as $r\rightarrow 0$ supporting the
Eqs.(\ref{Eq:HigherOrder}) and (\ref{Eq:Universal2}).

\end{widetext}

\section{Results: General $k$ Case}
\label{sec:ResultsKisGeneral}
In the case of general $k$, for Wigner-like surmise, we need to have $2k+1$ eigenvalues \cite{rao2020higher}.
Then the $k$-th order spacing ratio is defined as:
\begin{eqnarray}
 r^{(k)}=\frac{E_{2k+1}-E_{k+1}}{E_{k+1}-E_1}.
\end{eqnarray}
\begin{widetext}
Considering the Gaussian ensemble with $N=2k+1$ eigenvalues, the distribution of $r^{(k)}$ is given by:
%
\begin{eqnarray}
\label{Eq:GeneralKcase}
 P\left(r^{(k)}\right) \propto \int_{-\infty}^\infty \cdots \int_{-\infty}^\infty    \prod_{1\leq i<j\leq 2k+1}^{} |E_i-E_j|^{\beta} \;\;
\exp\left(-A \sum_{i=1}^{2k+1} E_i^2 \right) \delta\left(r^{(k)}-\dfrac{E_{2k+1}-E_{k+1}}{E_{k+1}-E_1}\right) \prod_{i=1}^{2k+1} dE_i.
\end{eqnarray}
After changing the variables as: $x_i=E_{i+1}-E_i$ for $i=1$ to $2k$ and $x_{2k+1}=\sum_{i=1}^{2k+1} E_i$ we get \cite{rao2020higher}:
\begin{eqnarray}
P\left(r^{(k)}\right) \propto  &&   \int_{0}^\infty \cdots \int_{0}^\infty \dfrac{\partial(E_1,\ldots,E_{2k+1})}{\partial(x_1,\ldots,x_{2k+1})}\nonumber
\left( \prod_{i=1}^{2k} \prod_{j=i}^{2k} \left|\sum_{l=i}^j x_l\right|^{\beta} \right)
\exp\left\{-\dfrac{A}{2k+1} \left[    \sum_{i=1}^{2k}\sum_{j=i}^{2k}\left(\sum_{l=i}^{j} x_l \right)^2 + x_{2k+1}^2\right] \right\}\\
&&\delta\left(r^{(k)}-\frac{\sum_{i=k+1}^{2k} x_i}{\sum_{i=1}^k x_i}\right) \prod_{i=1}^{2k+1} dx_i.
\end{eqnarray}
Here, the Jacobian $\dfrac{\partial(E_1,\ldots,E_{2k+1})}{\partial(x_1,\ldots,x_{2k+1})}$ and the integral for $x_{2k+1}$ are 
constants that can be absorbed into the normalization factor. Using the property of the delta function we obtain: 

\begin{eqnarray} \label{Eq:Integral4}
\begin{split}
P\left(r^{(k)}\right) \propto&   \int_{0}^\infty \cdots \int_{0}^\infty \left(\sum_{i=1}^k x_i\right)
\left( \prod_{i=1}^{2k} \prod_{j=i}^{2k} \left|\sum_{l=i}^j x_l\right|^{\beta} \right)
\exp\left\{-\dfrac{A}{2k+1} \left[\sum_{i=1}^{2k}\sum_{j=i}^{2k}\left(\sum_{l=i}^{j} x_l \right)^2 \right] \right\}\\
&\delta\left(r^{(k)}\sum_{i=1}^k x_i -{\sum_{i=k+1}^{2k} x_i} \right) \prod_{i=1}^{2k} dx_i.
\end{split}
\end{eqnarray}
Here, the integration is over $2k$ variables. First the
integration over $x_{2k}$ is carried out. In that case the delta function goes away replacing 
$x_{2k}$ by the following:
\begin{eqnarray}
\label{Eq:x2kEquation}
r^{(k)}=\frac{\sum_{i=k+1}^{2k} x_i}{\sum_{i=1}^k x_i}\implies x_{2k}=r^{(k)}\sum_{i=1}^k x_i-\sum_{i=k+1}^{2k-1} x_i.
\end{eqnarray}
This will put a constraint on the other variables $x_i$ for $i=k+1$ to $2k-1$, such that 
$0\leq \sum_{i=k+1}^{2k-1} x_i \leq r^{(k)}\sum_{i=1}^k x_i$. Thus, we need to find the polynomial that depends only on $x_i$ 
for all $i=k+1$ to $2k$. Following on the lines of previous sections we can write:

\begin{eqnarray}\label{Eq:productx2k1}
\prod_{i=1}^{2k} \prod_{j=i}^{2k} \left|\sum_{l=i}^j x_l\right|^{\beta} &=&
\prod_{i=1}^{k} \prod_{j=i}^{2k} \left|\sum_{l=i}^j x_l\right|^{\beta}  
\times \prod_{i=k+1}^{2k} \prod_{j=i}^{2k} \left|\sum_{l=i}^j x_l\right|^{\beta}  \\
&=&\label{Eq:productx2k2} \left| \prod_{i=1}^{k} \prod_{j=i}^{2k}  \left(\sum_{l=i}^j x_l\right ) \right|^{\beta}   \times 
\prod_{i=k+1}^{2k} \prod_{j=i}^{2k} \left|\sum_{l=i}^j x_l\right|^{\beta}  \\
&=&\label{Eq:productx2k3} \left({\tilde f}_1(x_1,\ldots,x_k)+{\tilde f}_2(x_1,\ldots,x_{2k})\right)^{\beta} \times 
\prod_{i=k+1}^{2k} \prod_{j=i}^{2k} \left|\sum_{l=i}^j x_l\right|^{\beta}.
\end{eqnarray}

In the first step we have splitted the product from $i=1$ to $2k$ in two terms such that the first one has the range
of $i=1$ to $k$ while second has $i=k+1$ to $2k$. This gives right hand side of the Eq.(\ref{Eq:productx2k1}). 
The first multinomial term in the Eq.(\ref{Eq:productx2k2}) is fully expanded such that it is sum of $\tilde f_1$ 
and $\tilde f_2$, where $\tilde f_1$ and $\tilde f_2$ are polynomial functions of the respective arguments only. 
The whole purpose of this split is to separate out terms containing the variables $x_1,\ldots,x_k$ only.
This is possible because every product term in 
$ \left| \prod_{i=1}^{k} \prod_{j=i}^{2k}  \left(\sum_{l=i}^j x_l\right ) \right|^{\beta} $ contains at least one 
variable from the set $\{x_1,\ldots,x_k\}$. This will show that when $(\tilde f_1+\tilde f_2)^{\beta}$ is fully expanded using 
binomial theorem, will imply that lowest degree of polynomial terms containing $x_{k+1},\ldots,x_{2k}$ is zero.
Thus, we get the following:
\begin{eqnarray} \label{Eq:Psplit1}
\begin{split}
P\left(r^{(k)}\right)\propto   \int_{0}^\infty \cdots \int_{0}^\infty & \left(\sum_{i=1}^k x_i\right) \times
\prod_{i=k+1}^{2k} \prod_{j=i}^{2k} \left|\sum_{l=i}^j x_l\right|^{\beta} \times
\left({\tilde f}_1(x_1,\ldots,x_k)+{\tilde f}_2(x_1,\ldots,x_{2k})\right)^{\beta}\\
& \times \exp\left\{-\dfrac{A}{2k+1} \left[\sum_{i=1}^{2k}\sum_{j=i}^{2k}\left(\sum_{l=i}^{j} x_l \right)^2 \right] \right\}
\delta\left(r^{(k)}\sum_{l=1}^k x_l -{\sum_{l=k+1}^{2k} x_l} \right) \prod_{i=1}^{2k} dx_i.
\end{split}
\end{eqnarray}
We will now further split the term $\prod_{i=k+1}^{2k} \prod_{j=i}^{2k} \left|\sum_{l=i}^j x_l\right|^{\beta} $
such that the terms containing $x_{2k}$ are separated out as follows:
\begin{eqnarray}\label{Eq:Psplit2}
 \prod_{i=k+1}^{2k} \prod_{j=i}^{2k} \left|\sum_{l=i}^j x_l\right|^{\beta}=
 \left(\prod_{i=k+1}^{2k-1} \prod_{j=i}^{2k-1} \left|\sum_{l=i}^j x_l\right|^{\beta} \right) 
\left(\prod_{i=k+1}^{2k} \left|\sum_{l=i}^{2k} x_l\right|^{\beta} \right). 
\end{eqnarray}
This is done because we will be first integrating over the variable $x_{2k}$. Thus, combinig Eqs.(\ref{Eq:Psplit1}) and 
(\ref{Eq:Psplit2}) we get:
\begin{eqnarray} \label{Eq:ProbRk}
\begin{split}
P\left(r^{(k)}\right)\propto   \int_{0}^\infty \cdots \int_{0}^\infty  & \left(\sum_{i=1}^k x_i\right) 
\left(\prod_{i=k+1}^{2k-1} \prod_{j=i}^{2k-1} \left|\sum_{l=i}^j x_l\right|^{\beta} \right) 
\left(\prod_{i=k+1}^{2k} \left|\sum_{l=i}^{2k} x_l\right|^{\beta} \right) 
\left({\tilde f}_1(x_1,\ldots,x_k)+{\tilde f}_2(x_1,\ldots,x_{2k})\right)^{\beta}\\
&\exp\left\{-\dfrac{A}{2k+1} \left[\sum_{i=1}^{2k}\sum_{j=i}^{2k}\left(\sum_{l=i}^{j} x_l \right)^2 \right] \right\}
\delta\left(r^{(k)}\sum_{l=1}^k x_l -{\sum_{l=k+1}^{2k} x_l} \right) \prod_{i=1}^{2k} dx_i.
\end{split}
\end{eqnarray}

Now, solving for the $x_{2k}$-integral will remove the delta function and replace $x_{2k}$ by 
$r^{(k)}\sum_{l=1}^k x_i-\sum_{l=k+1}^{2k-1} x_i$ at all the places in the integral as discussed in the Eq.(\ref{Eq:x2kEquation}).
First consider the term $ \prod_{i=k+1}^{2k} \left|\sum_{l=i}^{2k} x_l\right|^{\beta} $ from the Eq.(\ref{Eq:ProbRk}).
It can be written and simplified further using $x_{2k}=r^{}\sum_{l=1}^k x_l-\sum_{l=k+1}^{2k-1} x_l$ as follows (here,
$r^{(k)}=r$ is defined for simplicity of the notation):
\begin{eqnarray}
\prod_{i=k+1}^{2k} \left|\sum_{l=i}^{2k} x_l\right|^{\beta} &=& \left|\sum_{l=k+1}^{2k} x_l\right|^{\beta}  \times
\prod_{i=k+2}^{2k} \left|\sum_{l=i}^{2k} x_l\right|^{\beta}\\
&=&\label{Eq:simplify22} \left(r\sum_{l=1}^{k}x_l\right)^{\beta} \times
\left(\prod_{i=k+2}^{2k} \left(r\sum_{l=1}^{k}x_l-\sum_{l=k+1}^{i-1} x_l\right)\right)^{\beta}.
\end{eqnarray}
Similarly, using the constraint in Eq.(\ref{Eq:x2kEquation}) one obtains:
\begin{eqnarray}\label{Eq:f1f2simple1}
\begin{split}
\left({\tilde f}_1(x_1,\ldots,x_k)+{\tilde f}_2(x_1,\ldots,x_{2k})\right)^{\beta} &\rightarrow
\left({\tilde f}_1\left(x_1,\ldots,x_k\right)+{\tilde f}_2\left(x_1,\ldots,r^{(k)}\sum_{l=1}^k x_i-\sum_{l=k+1}^{2k-1} x_i\right)\right)^{\beta}\\
\mbox{and}\;\;\;\;\; \sum_{i=1}^{2k}\sum_{j=i}^{2k}\left(\sum_{l=i}^{j} x_l \right)^2 &\rightarrow \sum_{j>i=1}^{2k-1} x_i x_j h_{ij}'(r),
\end{split}
\end{eqnarray}
where $h_{ij}'$ are polynomials in $r$.
%
%
%
%
Since $x_i \geq 0$ for all $i$, in order to have all integrals converging it is sufficient to show that $h_{ii}' > 0$ for all $i$.
This will be shown now. Considering the following term from the exponent of the Eq.(\ref{Eq:ProbRk}) and simplifying it we get
(see the text followed for the steps on the simplifications done at each stage):
\begin{eqnarray}
 \sum_{i=1}^{2k} \sum_{j=i}^{2k} \left(\sum_{l=i}^{j} x_l \right)^2&=& \label{Eq:sum1}
 \sum_{i=1}^{2k-1} \sum_{j=i}^{2k-1}  \left(\sum_{l=i}^{j} x_l \right)^2     +
 \sum_{i=1}^{2k-1} \left(\sum_{l=i}^{2k} x_l \right)^2  + \left(   x_k \right)^2\\
&=& \label{Eq:sum2} \sum_{i=1}^{2k-1} \sum_{j=i}^{2k-1}  \left(\sum_{l=i}^{j} x_l \right)^2 +
 \sum_{i=1}^{k} \left(\sum_{l=i}^{2k} x_l \right)^2 +  \left(\sum_{l=k+1}^{2k} x_l \right)^2+
 \sum_{i=k+2}^{2k-1} \left(\sum_{l=i}^{2k} x_l \right)^2 +\left(   x_k \right)^2\\
 &=& \label{Eq:sum3} \sum_{i=1}^{2k-1} \sum_{j=i}^{2k-1}  \left(\sum_{l=i}^{j} x_l \right)^2 +
 \sum_{i=1}^{k} \left(\sum_{l=i}^{k} x_l  +\sum_{l=k+1}^{2k} x_l  \right)^2 +  \left(\sum_{l=k+1}^{2k} x_l \right)^2+\\
&&  \sum_{i=k+2}^{2k-1} \left(  \sum_{l=k+1}^{2k} x_l  - \sum_{l=k+1}^{i-1} x_l \right)^2 +\left(   x_k \right)^2 \nonumber\\
&=& \label{Eq:sum4}  \sum_{i=1}^{2k-1} \sum_{j=i}^{2k-1}  \left(\sum_{l=i}^{j} x_l \right)^2 +
 \sum_{i=1}^{k} \left(\sum_{l=i}^{k} x_l  + r \sum_{l=1}^{k} x_l  \right)^2 +  \left( r \sum_{l=1}^{k} x_l  \right)^2+\\
&& \sum_{i=k+2}^{2k-1} \left(  r \sum_{l=1}^{k} x_l  - \sum_{l=k+1}^{i-1} x_l \right)^2 +\left(   x_k \right)^2. \nonumber 
\end{eqnarray}
Here, the Eq.(\ref{Eq:sum1}) is obtained by splitting the summation such that term $x_{2k}$ is separated out.
The summation $i=1$ to $2k-1$  in the second term of Eq.(\ref{Eq:sum1}) is further splitted into three parts: 
summation $i=1$ to $k$, single term $i=k+1$, and summation $i=k+2$ to $2k-1$ to get the Eq.(\ref{Eq:sum2}).
The summation $l=i$ to $2k$ in the second and fourth term in Eq.(\ref{Eq:sum2}) is further spitted depending on the range of 
$i$, so that we can use the constraint from Eq.(\ref{Eq:x2kEquation}). This will give us the Eq.(\ref{Eq:sum3}).
The Eq.(\ref{Eq:sum4}) is obtained by using the same constraint in the Eq.(\ref{Eq:sum3}). We can see from the Eq.(\ref{Eq:sum4}) 
that after exapansion each term in the coefficient of $x_i^2$ (for all $i$) is either a positive number (at least one such number 
exists and is ensured by the first term in the Eq.(\ref{Eq:sum4})) or a function of $r$ and can be seen to be always 
non-negative. 
The only terms with negative sign come from the second last term in the Eq.~\ref{Eq:sum4}, which only contains mixed
terms like $x_ix_j$ with $i \neq j$.
Denoting the coefficient of $x_i x_j$ by $h_{ij}'$ we have  $h_{ii}'>0$ for all $i$, thus proving our claim. 
The exact expressions 
for $h_{ij}'$ is not required for our purpose here.
%
%
Thus, the $h_{ii}'$ are polynomials in $r$ such that in the limit $r\rightarrow0$ they are all non-zero, which makes the integral
converging. Thus, combining the Eqs.~\ref{Eq:x2kEquation}, \ref{Eq:ProbRk}, \ref{Eq:simplify22} and \ref{Eq:f1f2simple1} we get:
\begin{eqnarray} \label{ }
\begin{split}
P\left(r^{(k)}\right)\propto & \idotsint \limits_{x_1,\ldots,x_k=0}^{\infty}\;\;\;\;\;\idotsint \limits_
{0\leq \sum_{i=k+1}^{2k-1} x_i\leq r\left(\sum_{i=1}^{k} x_i\right)}  \left(\sum_{i=1}^{k}x_i\right) 
\left(\prod_{i=k+1}^{2k-1} \prod_{j=i}^{2k-1} \left|\sum_{l=i}^j x_l\right|^{\beta} \right)
\left(r\sum_{l=1}^{k}x_l\right)^{\beta}\\
&\left(\prod_{i=k+2}^{2k} \left(r\sum_{l=1}^{k}x_l-\sum_{l=k+1}^{i-1} x_l\right)\right)^{\beta}
\left({\tilde f}_1\left(x_1,\ldots,x_k\right)+{\tilde f}_2\left(x_1,\ldots,x_{2k-1},r\sum_{l=1}^{k}x_l-\sum_{l=k+1}^{2k-1}x_l\right)\right)^{\beta}\\
&\exp\left\{-\dfrac{A}{2k+1} \left[\sum_{j>i=1}^{2k-1} x_i x_j h_{ij}'(r) \right] \right\} \;\prod_{i=1}^{2k-1} dx_i.
\end{split}
\end{eqnarray}

Next we rewrite the integral such that the summation term in the exponential term gets divided into parts. One part contains variables 
only from $x_1$ to $x_k$ and the other term containing all of them i.e. from $x_1$ to $x_{2k-1}$.
Thus, we get:
\begin{eqnarray} \label{Eq:PrkI1}
\begin{split}
P\left(r^{(k)}\right)\propto & \idotsint \limits_{x_1,\ldots,x_k=0}^{\infty}  I_{x_{k+1}\ldots x_{2k-1}} 
\left(\sum_{l=1}^{k}x_l\right)^{1+\beta}
\exp\left\{-\dfrac{A}{2k+1} \left[\sum_{j>i=1}^{k} x_i x_j h_{ij}'(r) \right] \right\} \;\prod_{i=1}^{k} dx_i,
\end{split}
\end{eqnarray}
where
\begin{eqnarray} \label{Eq:Integral50}
\begin{split}
I_{x_{k+1}\ldots x_{2k-1}} = &\idotsint \limits_{0\leq \sum_{i=k+1}^{2k-1} x_i\leq r\left(\sum_{i=1}^{k} x_i\right)} r^{\beta}
\left(\prod_{i=k+1}^{2k-1} \prod_{j=i}^{2k-1} \left|\sum_{l=i}^j x_l\right|^{\beta} \right)
\left(\prod_{i=k+2}^{2k} \left(r\sum_{l=1}^{k}x_l-\sum_{l=k+1}^{i-1} x_l\right)^{\beta}\right)\\
&\left({\tilde f}_1\left(x_1,\ldots,x_k\right)+{\tilde f}_2\left(x_1,\ldots,x_{2k-1},r\sum_{l=1}^{k}x_l-\sum_{l=k+1}^{2k-1}x_l\right)\right)^{\beta}\\
&\exp\left\{-\dfrac{A}{2k+1} \left[\sum_{i=1,j=k+1}^{2k-1} x_i x_j h_{ij}'(r) \right] \right\} \;\prod_{i=k+1}^{2k-1} dx_i.
\end{split}
\end{eqnarray}
It can be seen from the Eq.(\ref{Eq:PrkI1}) that in the limit $r\rightarrow 0$ the leading order of $r$ will only come from evaluating 
that for $I_{x_{k+1}\ldots x_{2k-1}}$. In the subsequent part of the paper we will derive the latter. 
Now, consider the term $\prod_{i=k+2}^{2k} \left(r\sum_{l=1}^{k}x_l-\sum_{l=k+1}^{i-1} x_l\right)^{\beta}$ from the 
Eq.(\ref{Eq:Integral50}). This can be simplified as follows (assuming that $\beta$ is a natural number):
\begin{eqnarray} \label{ }
\begin{split}
\prod_{i=k+2}^{2k} \left(r\sum_{l=1}^{k}x_l-\sum_{l=k+1}^{i-1} x_l\right)^{\beta}=&
\prod_{i=k+2}^{2k} \sum_{q=0}^{\beta}{\beta \choose q} \left(r\sum_{l=1}^{k}x_l\right)^{q} \left( -\sum_{l=k+1}^{i-1} x_l\right)^{\beta-q}\\
=& \prod_{i=k+2}^{2k} \sum_{q=0}^{\beta}{\beta \choose q} \left(\sum_{l=1}^{k}x_l\right)^{q} r^q \left( -\sum_{l=k+1}^{i-1} x_l\right)^{\beta-q}.
\end{split} 
\end{eqnarray}
Thus, the Eq. (\ref{Eq:Integral50}) simplifies to:
\begin{eqnarray} \label{Eq:Ixk1}
\begin{split}
&I_{x_{k+1}\ldots x_{2k-1}} = \idotsint \limits_{0\leq \sum_{i=k+1}^{2k-1} x_i\leq r\left(\sum_{i=1}^{k} x_i\right)} 
\left[ r^{\beta}\right] \times
\left[ \prod_{i=k+1}^{2k-1} \prod_{j=i}^{2k-1} \left|\sum_{l=i}^j x_l\right|^{\beta} \right] \\
&\left[\prod_{i=k+2}^{2k} \sum_{q=0}^{\beta}{\beta \choose q} \left(\sum_{l=1}^{k}x_l\right)^{q} r^q 
\left( -\sum_{l=k+1}^{i-1} x_l\right)^{\beta-q}
\right]
\left[{\tilde f}_1\left(x_1,\ldots,x_k\right)+{\tilde f}_2\left(x_1,\ldots,x_{2k-1},r\sum_{l=1}^{k}x_l-\sum_{l=k+1}^{2k-1}x_l\right)\right]^{\beta}\\
&\exp\left\{-\dfrac{A}{2k+1} \left[\sum_{i=1,j=k+1}^{2k-1} x_i x_j h_{ij}'(r) \right] \right\} \;\prod_{i=k+1}^{2k-1} dx_i.
\end{split}
\end{eqnarray}
The square brackets around various terms are put in order to address them individually.
Here, we will be using the following integral identity (the generalization of the Eq.~(\ref{Eq:IntegralIdentity1})):
\begin{eqnarray}\label{Eq:Integral5}
  \idotsint\limits_{0\leq y_1,\ldots, y_N, \sum_{i=1}^{N} y_i \leq a}  \prod_{i=1}^{N} y_i^{p_i} dy_i \;  \propto \; a^{\sum_{i=1}^{N}p_i +N}.
\end{eqnarray}
In Eq.~(\ref{Eq:Integral5}), it should be noted that the exponent on the right-hand side is a function only of the order of 
the integrand polynomial $\left(\sum_{i=1}^{N}p_i\right)$ and the number of variables ($N$) on the left-hand side.
Here, we are interested only in the limit $r\rightarrow 0 $. 
Thus, we need to find the lowest order of $r$ in 
$I_{x_{k+1}\ldots x_{2k-1}}$. For that we need to first find the lowest order of the polynomial in $x_{k+1}$ to $x_{2k-1}$ in 
Eq.~(\ref{Eq:Ixk1})
and then use Eq.~(\ref{Eq:Integral5}). This can be achieved by doing the same for each term in the Eq.~(\ref{Eq:Ixk1}),  
multiplying them together, and then use the Eq.~(\ref{Eq:Integral5}). This is now explained in the next paragraph.

The first square bracket in the Eq.(\ref{Eq:Ixk1}) will give us an exponent of $\beta$ for $r$. The term 
$\prod_{i=k+1}^{2k-1} \prod_{j=i}^{2k-1} \left|\sum_{l=i}^j x_l\right|^{\beta}$ from the second bracket
 is a multinomial term and can be expanded fully.   It will lead to a homogeneous polynomial of degree $(k-1)k\beta/2$.
%
%
%
In the third square bracket, the term $\left(\sum_{l=1}^{k}x_l\right)^{q}$ do not have any of the variables from the set 
$\{ x_{k+1},\ldots, x_{2k}\}$. Thus, it is not going to give any $r$-dependent factor in the limit $r \rightarrow 0$. 
Thus, we are left with two terms,  namely $r^q$ and $ \left( -\sum_{l=k+1}^{i-1} x_l\right)^{\beta-q}$. 
Here, it can be seen that the term $\left( -\sum_{l=k+1}^{i-1} x_l\right)^{\beta-q}$ when expanded will give a 
homogeneous polynomial of order $\beta-q$. Both of them appears $(k-1)$ 
times due to the operation $\prod_{i=k+2}^{2k}$ on them. 
The range of the summation in $\left( -\sum_{l=k+1}^{i-1} x_l\right)^{\beta-q}$ do change with $i$
but the order of the homogeneous polynomial remains same.
Thus, using Eq.~(\ref{Eq:Integral5}) and 
$r\rightarrow0$ we can say that the third square bracket will result in an exponent of $ (k-1)q+(k-1)(\beta-q)$. 
The exponent of the lowest order polynomial in $x_{k+1},\ldots,x_{2k-1}$ which can be obtained from the term
in the fourth square bracket, namely $({\tilde f}_1 +{\tilde f}_2)^{\beta}$ is $0$.
This is because ${\tilde f}_1$ is a function of $x_1\ldots x_k$ {\it only} and use of binomial theorem (assuming $\beta$ is natural 
number) we get at least one term with variables $x_1\ldots x_k$ only.
It means that the lowest order of the polynomial containing $x_{k+1}\ldots x_{2k}$ variables will be zero.
While that from the exponential term (fifth term), using its Taylor's expansion, is also $0$.
%
%
Finally, the integration measure $\prod_{i=k+1}^{2k-1} dx_i$ has $k-1$ variables. Thus, the exponent of 
$r=\beta'$ where  
$\beta'=\left[ \beta\right]+ \left[k(k-1)\beta/2\right]+ \left[(k-1)q+(k-1)(\beta-q)\right]+ \left[0\right]+ \left[0\right]+ 
\left[(k-1)\right]=\beta k(k+1)/2+k-1$.

Now, using the identity from Eq.~(\ref{Eq:Integral5}) it can be seen that in the limit $r\rightarrow0$ the dominant term 
will be proportional to $r^{\beta'}$ where $\beta'=\beta+k(k-1)\beta/2+(k-1)(q+(\beta-q))+(k-1)=\beta k(k+1)/2+k-1$.
Thus, the leading term in $I_{x_{k+1}\ldots x_{2k-1}}$ in the limit $r \rightarrow 0$ is $r^{\beta'}$
which will also be the same for $P\left(r^{(k)}\right)$ as discussed earlier. Thus, we can write:
\begin{eqnarray}\label{Eq:Rtendstozero}
 P\left(r^{(k)}\right) \rightarrow \left(r^{(k)}\right)^{\beta'} \;\;\mbox{for}\;\; {r^{(k)}\to 0}.
\end{eqnarray}
With this, we have proved first part of the most general and main result in the Eq.~(\ref{Eq:Universal2})
supporting the Eq.~(\ref{Eq:HigherOrder}).


\end{widetext}

\section{Case of $r\rightarrow\infty$}
\label{Sec:ResultsRinInfinity}
In order to find the limiting behaviour in this case we use the property of the jpdf in the Eq.~(\ref{Eq:jpdfGaussian}). For this 
we show that $P\left(s_1,s_2\right)=P\left(s_2,s_1\right)$ i.e. $P\left(s_1,s_2\right)$ is a symmetric function, where 
$s_1=E_{k+1}-E_{1}$, $s_2=E_{2k+1}-E_{k+1}$, $P\left(s_1,s_2\right)$ is a jpdf of $s_1$ and $s_2$. We will show this for the 
Wigner-surmise setting, as per the Eq.~(\ref{Eq:GeneralKcase}), i.e for given $k$ we have $N=2k+1$.
Using the change of variables as per Sec.\ref{sec:ResultsKisGeneral} we get $s_1=\sum_{i=1}^k x_i$ and $s_2=\sum_{i=k+1}^{2k} x_i$
\cite{rao2020higher}. Now, using a property of the jpdf in the Eq.~(\ref{Eq:jpdfGaussian}) it can be seen that, it is invariant 
under the transformation: $x_i\leftrightarrow x_{2k+1-i}$, where $i=1$ to $k$. This corresponds to a reflection symmetry about the
eigenvalue $E_{k+1}$. It results in $s_1 \leftrightarrow s_2$. Thus, the jpdf is invariant so is the $P\left(s_1,s_2\right)$ under 
the said transformation i.e. 
$P\left(s_1,s_2\right)=P\left(s_2,s_1\right)$. Due to this left-right symmetry the distribution of $r^{(k)}=s_1/s_2$ is same as that 
of $1/r^{(k)}$ so that the following duality relation holds true:
\begin{eqnarray}
P\left(r^{(k)}\right)= \dfrac{1}{(r^{(k)})^2 } P\left(\dfrac{1}{r^{(k)}}\right),
\end{eqnarray}
where $P\left( x \right)$ is the probability distribution of $x$. The same relation corresponding to $k=1$ was presented earlier in 
the Ref.\cite{BogomolnyDistribution2013}. Thus, we can find the asymptotic behaviour of $r\rightarrow \infty$ using the solved 
case of $r\rightarrow 0$ in the Eq.~(\ref{Eq:Rtendstozero}). Thus, 
\begin{eqnarray}
\begin{split}
\lim_{r^{(k)}\to\infty} P(r)&= \lim_{r^{(k)}\to\infty} \;\; \dfrac{1}{(r^{(k)})^2 } P\left(\dfrac{1}{r^{(k)}}\right)\\
&=   \lim_{t\to 0} \;\;  t^2  P(t)\;\;\; \mbox{where}\;\;\; t=\dfrac{1}{r^{(k)}} \\
&= t^{2+\beta'}\\
&= (r^{(k)})^{-2-\beta'}.
\end{split}
\end{eqnarray}
Thus, we get the following result:
\begin{eqnarray}\label{Eq:RtendstoInfinity}
P\left(r^{(k)}\right)  \rightarrow \left(r^{(k)}\right)^{-2-\beta'} \;\; \mbox{for} \;\; \; r^{(k)}\to \infty.
%
\end{eqnarray}
With this, the second part of the Eq.~(\ref{Eq:Universal2}) is proved. It must be noted that we have shown the $r\rightarrow\infty$ 
behavior using the Wigner-like surmise i.e. for given order $k$ matrix dimension is $2k+1$. For cases otherwise, the symmetry of 
$P\left(s_1,s_2\right)$ holds only in the bulk of the spectrum and in the limit $N\rightarrow\infty$.
This symmetry will break down at the soft or hard edge of the spectrum, and deviations can be expected.

\section{Case of uncorrelated spectra}
\label{Sec:UncorrelatedSpectra}
Let's now consider the case of uncorrelated spectra. NN spacing ratio of such spectra shows Poissonian behaviour 
which is shown by integrable systems \cite{Oganesyan2007,Atas2013}. Higher-order spacing ratios, in this case, are known as 
follows \cite{tekur2018symmetry}:
\begin{eqnarray}
P_{P}^{k}(r) =\dfrac{(2k-1)!}{((k-1)!)^2} \dfrac{r^{k-1}}{(1+r)^{2k}}.
\end{eqnarray}
Important to note that this is an exact result in the limit of $N\rightarrow\infty$ only,
in contrast to many other equations in this paper. It can be shown easily that 
\begin{eqnarray}
P_{P}^{k}(r) \rightarrow r^{k-1} \;\;\;\; \mbox{for}\;\;  \; r^{}\to 0
\end{eqnarray}
and
\begin{eqnarray}
P_{P}^{k}(r) \rightarrow r^{-k-1} \;\;\;\; \mbox{for}\;\;  \; r^{}\to \infty.
\end{eqnarray}
This is a special case of our result above for $\beta'$ evaluated at $\beta=0$.

%

\section{SUMMARY AND CONCLUSIONS}
\label{Sec:Summary}

In recent times, higher-order spacing ratios have become a popular and important measure to study fluctuations in random matrices
and complex physical systems. This is due to their computationally simple nature as no unfolding is required,
compared to that of the {\it spacings} alone. Very few analytical results for the spacing ratios are available. 
This paper has analytically studied the asymptotic behaviour of higher-order spacing ratios ($r^{(k)}$) in the Gaussian ensembles
with Dyson index $\beta$. Most of the results on it were numerical 
\cite{UdaysinhScaling2018,tekurhigher2018,rao2020higher,rao2022random,PiotrSierantLevelStat2019,gong2020comparison}.
We have now proved an universal behavior of its distribution i.e. $P^{k}(r,\beta)  \rightarrow r^{\beta'}$ ($ r^{-2-\beta'}$)
in the limit $r \rightarrow 0 $ ($ \infty $), where $\beta'=\beta k(k+1)/2+ (k-1)$ based on the very good
approximate Eq.(\ref{Eq:HigherOrder}).
We also expect the same behavior by the exact expression (currently unknown) for $P^{k}(r,\beta)$.
We have used the Wigner-like surmise (Eq.(\ref{Eq:HigherOrder})) which becomes a good fit for the large-$N$ scenario.
%
Here, universality is refered to in the sense that the ratios can be studied without the procedure of unfolding or 
renormalization of the eigenvalues which is very much required in the case of the spacings \cite{Mehtabook,Forresterbook}.
In fact, from our study of uncorrelated eigenvalues, our results hold true for any $\beta \geq 0$.
These results have given analytical support to the numerical results from various random matrix ensembles and complex physical 
systems, which was absent earlier 
\cite{UdaysinhScaling2018,tekurhigher2018,rao2020higher,rao2022random,PiotrSierantLevelStat2019,gong2020comparison}.
Moreover, our analytical approach can be extended to other ensembles, for example Laguerre ensemble 
\cite{Forresterbook,akemann2011oxford,Wishart28,Satyalargeright,Fridman12,UdaysinhBhosaleScaling2018}, chiral ensembles 
\cite{verbaarschot1994spectrum,Verbaarschot94,verbaarschot2000random,fyodorov2002correlation,damgaard2011chiral,kaymak2014supersymmetry,BeenakkerMajorana2015,10.1093/oso/9780198797319.003.0005,PragyaShuklaChiral2020,RichterMicrowave2020},
etc. Though Laguerre and chiral ensembles are related to each other mathematically they have different applications. Wishart 
ensembles are used in the study of entanglement \cite{akemann2011oxford}, wireless communication systems \cite{Forresterbook} whereas,
chiral ensembles are used to model Dirac operators in quantum chromodynamics
\cite{verbaarschot1994spectrum,Verbaarschot94,verbaarschot2000random}. Recently it is shown that the NN level spacings distribution 
is insensitive to the position in RMT spectra at the edges or in the bulk
despite the fact that fluctuations there are described by different limiting kernals \cite{akemann2022consecutive}.
We would like to investigate the same with the spacing ratios numerically as well as analytically.

It should be noted that we have given the asymptotic behaviour of higher-order spacing ratios but finding
an exact expression for the corresponding Wigner-like surmise still remains open. This is left for a future study.

\section{Acknowledgments}
The author is thankful to M. S. Santhanam, Harshini Tekur, Ravi Prakash and Harshit Sharma for valuable comments and discussions at 
various levels of this paper.  

 
 

\bibliography{reference22013,reference22,reference221}
\end{document}